\documentclass{article}
\makeatletter
\@addtoreset{equation}{section}

\makeatother

\usepackage{graphics}

\begin{document}

\begin{flushright}
Jun 2007

KUNS-2078
\end{flushright}

\begin{center}

\vspace{3cm}

{\LARGE 
\begin{center}
$AdS$/CFT Correspondence

as 

A Consequence of Scale Invariance
\end{center}
}

\vspace{2cm}

Hikaru Kawai $^{ab}$\footnote{e-mail address : hkawai@gauge.scphys.kyoto-u.ac.jp} \hspace{2mm}
and 
\hspace{2mm}
Takao Suyama $^a$\footnote{e-mail address : suyama@gauge.scphys.kyoto-u.ac.jp}

\vspace{1cm}

$^a$ {\it Department of Physics, Kyoto University,}

{\it Kitashirakawa, Kyoto 606-8502, Japan }

{\it and}

$^b$ {\it Theoretical Physics Laboratory,} 

{\it The Institute of Physics and Chemical Research (RIKEN),}

{\it Wako, Saitama 351-0198, Japan}

\vspace{3cm}

{\bf Abstract} 

\end{center}

We study an anisotropic 
scale transformation in the worldsheet description of D3-branes in Type IIB theory, and 
show that the transformation is really a symmetry in a region near D3-branes. 
$AdS$/CFT correspondence follows from this symmetry. 
We will explicitly show that 
Wilson loops in ${\cal N}=4$ supersymmetric Yang-Mills 
theory and minimal surfaces in $AdS_5$ are related by the symmetry.  
The functional form of a supersymmetric Wilson loop operator is naturally derived from our 
worldsheet point of view.

\newpage

\vspace{1cm}

\section{Introduction}

\vspace{5mm}

Since Maldacena proposed the conjectured relation, known as $AdS$/CFT correspondence 
\cite{Maldacena}, between 
${\cal N}=4$ supersymmetric Yang-Mills theory (SYM) and Type IIB string theory on $AdS_5\times S^5$ 
background, a vast amount of researches have been reported to provide pieces of supporting evidence 
for the correspondence as many as possible. 
Many of such researches discuss some limiting situations, for example, in the large $N$ limit, where 
$N$ is the rank of the gauge group. 
On the expected range of validity, however, there does not seem to be a consensus. 
The most well-studied limit is to take the limit of 
both the large $N$ and the large 't Hooft coupling $\lambda$. 
The correspondence claimed for this limit is sometimes referred to as the weakest version of $AdS$/CFT 
correspondence. 

One of the claims of the correspondence which is now well-established is the correspondence between 
Wilson loops 
$W(C)$ in ${\cal N}=4$ SYM and minimal surfaces $\Sigma$ in the $AdS_5$ background 
\cite{RY}\cite{Maldacena2}. 
The expectation value of $W(C)$ is related to the partition function of a single Type IIB string whose 
boundary $\partial \Sigma$ lies on $C$ embedded in the boundary of $AdS_5$. 
A similar correspondence was already anticipated in old days, 
and $AdS$/CFT correspondence provides a concrete example which can be 
verified quantitatively. 

In this paper, we would like to provide a natural point of view for understanding {\it why} such a 
correspondence holds. 
It is based on a perturbative worldsheet description of D3-branes in the flat space-time. 
We consider an anisotropic scale transformation 
\begin{equation}
x^\mu \to cx^\mu, \hspace{5mm} y^I \to c^{-1}y^I, 
\end{equation}
where $x^\mu$ is the longitudinal coordinates and $y^I$ are the transverse coordinates of the 
D3-branes. 
For simplicity, we call this the scale transformation. 
We will show that the correspondence between the Wilson loops and the minimal surfaces 
follows from the invariance under the scale transformation, which we call in this paper 
the scale invariance. 
In a region near D3-branes, this invariance can be realized 
in the worldsheet description, as far as closed string loops are suppressed. 
The existence of the scale invariance immediately implies, for example, 
the weakest version of the correspondence. 

This paper is organized as follows. 
In section \ref{sketch}, 
we provide an argument which indicates the importance of the scale invariance 
in $AdS$/CFT correspondence. 
Section \ref{boundary} is devoted to the review of boundary states in the Green-Schwarz formalism 
\cite{GG}. 
Since the space-time symmetries play crucial roles, and also the R-R states are important, 
it is convenient to employ 
the Green-Schwarz formalism. 
In section \ref{scale}, 
we define the scale transformation of the worldsheet variables, and show that the 
free energy of Type IIB string in the presence of D3-branes is invariant under the transformation 
in a region near D3-branes. 
In our worldsheet description, the functional form of a supersymmetric Wilson loop, 
which is studied in \cite{ESZ}\cite{DG}, is naturally 
derived without the detailed knowledge of the supersymmetry transformation, as shown in section 
\ref{Zarembo}. 
Section \ref{discuss} is devoted to discussion.

\vspace{1cm}

\section{Scale invariance in $AdS$/CFT} \label{sketch}

\vspace{5mm}

Let us recall the metric of the solution of Type IIB supergravity for $N$ D3-branes 
\begin{equation}
ds^2 = \left( 1+\frac{4\pi\lambda}{y^4} \right)^{-\frac12}\eta_{\mu\nu}dx^\mu dx^\nu
 +\left( 1+\frac{4\pi\lambda}{y^4} \right)^{\frac12}dy^Idy^I. 
    \label{D3soln}
\end{equation}
where $\mu$ and $\nu$ run from 0 to 3, and $I$ runs from 4 to 9. 
Here 
\begin{equation}
\lambda = g_sN 
\end{equation}
is the 't Hooft coupling. 
We have defined $y^2=y^Iy^I$. 
The string length $l_s$ has chosen to be unity. 
In a region $y^4<<\lambda$, which we call the near-horizon region, the metric behaves as 
\begin{equation}
ds^2 \to \sqrt{4\pi \lambda}\left[ y^2\eta_{\mu\nu}dx^\mu dx^\nu+\frac{dy^2}{y^2} \right]
 +\sqrt{4\pi \lambda}d\Omega_5^2, 
    \label{AdS}
\end{equation}
after the rescaling $y\to \sqrt{4\pi\lambda}y$. 
This is the metric on the direct product of $AdS_5$, in the Poincar\'e coordinates, and $S^5$. 
It is easily recognized that the scale transformation 
\begin{equation}
\delta_\epsilon x^\mu = -\epsilon x^\mu, \hspace{5mm} 
\delta_\epsilon y^I = \epsilon y^I, 
   \label{isometry}
\end{equation}
is an isometry of the metric (\ref{AdS}). 
This transformation shrinks the D3-branes in the longitudinal directions $x^\mu$ and expands 
in the transverse directions $y^I$. 
This is actually an $SO(1,1)$ subgroup of the isometry $SO(4,2)$ of $AdS_5$. 

\vspace{5mm}

Let us examine the transformation property of the original metric (\ref{D3soln}) 
under (\ref{isometry}). 
We obtain 
\begin{equation}
\delta_\epsilon (ds^2) = \epsilon \frac{y^4}{\pi\lambda}\cdot\lambda\partial_\lambda (ds^2). 
   \label{deltaG}
\end{equation}
For a fixed $\lambda$, the variation of the metric 
under the transformation (\ref{isometry}) is negligible in the near-horizon region. 
One of the implications of the existence of the near-horizon region is as follows. 
Consider an observable ${\cal O}[g]$ which depends only on the near-horizon region 
of the metric (\ref{D3soln}). 
The scale transformation of ${\cal O}[g]$ is negligible 
in the near-horizon region, due to (\ref{deltaG}), and therefore ${\cal O}[g]$ is invariant there. 
In this sense, the scale invariance appears in the near-horizon region. 

We would like to extend the above argument on classical gravity to string theory. 
We will define in section \ref{scale} 
the scale transformation (\ref{isometry}) for worldsheet variables, and examine the invariance of 
the total system of strings and D3-branes in a similar sense discussed above. 
In the rest of this section, we discuss implications of the existence of the symmetry in string 
theory. 

\begin{figure}[tbp]
\includegraphics{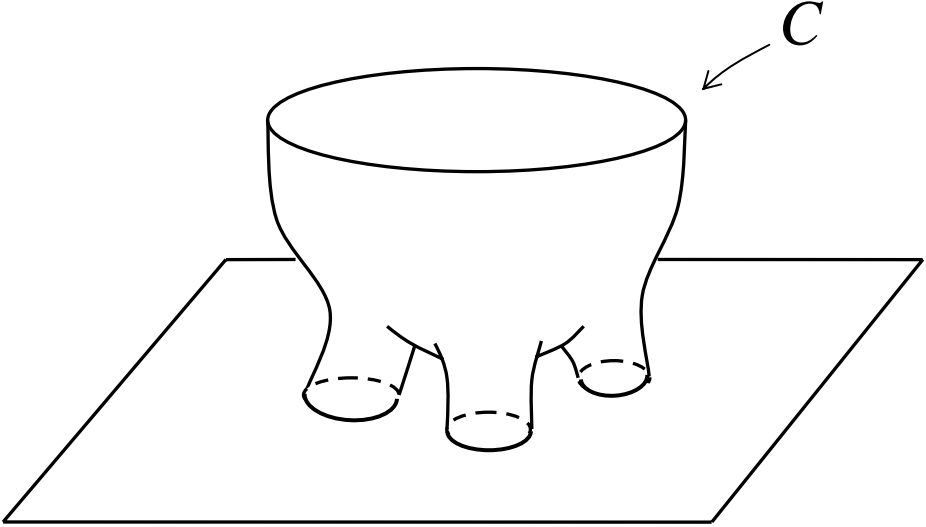}
\caption{A worldsheet with four boundaries. Three of them are on the D3-branes. The remaining one 
of them, denoted as $C$, is attached to a source, say another D-brane, apart from the D3-branes. }
   \label{intermediate}
\end{figure}

\vspace{5mm}

To be concrete, let us consider a system of 
a string and $N$ D3-branes, depicted in figure \ref{intermediate}. 
The string has a boundary $C$ which is away from the D3-branes. 
One can imagine that there is another D-brane on which $C$ lies. 
The worldsheet of the string may have the other boundaries which are on the D3-branes. 
We take $C$ to be within the near-horizon region, in order for the system to have the 
scale invariance. 

\begin{figure}[tbp]
\includegraphics{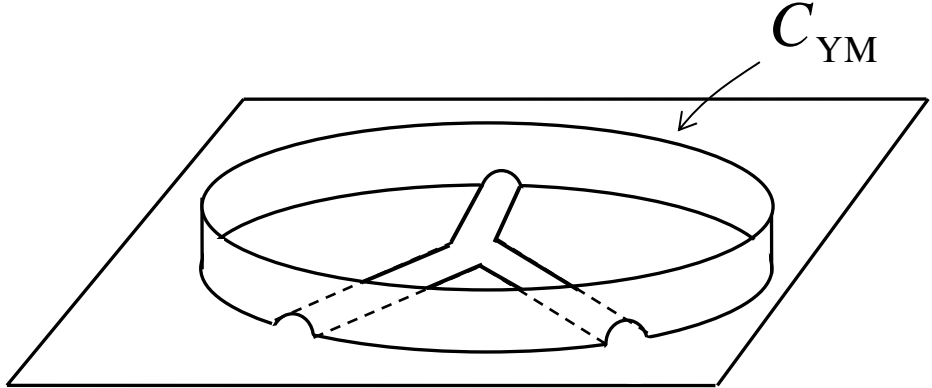}
\caption{A worldsheet near D3-branes. Open strings propagate inside the outer cylindrical 
worldsheet. }
   \label{gauge}
\end{figure}

On the one hand, we can expand the system in 
the $x^\mu$-directions and shrink in the $y^I$-directions 
by the scale transformation (\ref{isometry}). 
Then, the worldsheet will have a configuration depicted in figure \ref{gauge}. 
It is interesting to notice that this worldsheet configuration can be regarded as a Feynman diagram 
for a Wilson loop $W(C_{\rm YM})$ for a loop $C_{\rm YM}$ in the double-line notation. 
A worldsheet with more boundaries on the D3-branes corresponds to a more complicated Feynman diagram. 
The size of $C_{\rm YM}$ can be taken to be as large as we want, compared with the string scale. 
Therefore, the open strings propagating on the D3-branes have small momenta, 
and only the contributions 
from the massless states of the open strings survive. 
These arguments imply that the summation of the worldsheets with a fixed boundary $C_{\rm YM}$ 
in the presence of the D3-brane 
background actually evaluates a Wilson loop in ${\cal N}=4$ SYM in four-dimensions. 
We discuss the functional form of the Wilson loop in more details in section \ref{Zarembo}. 

\begin{figure}[tbp]
\includegraphics{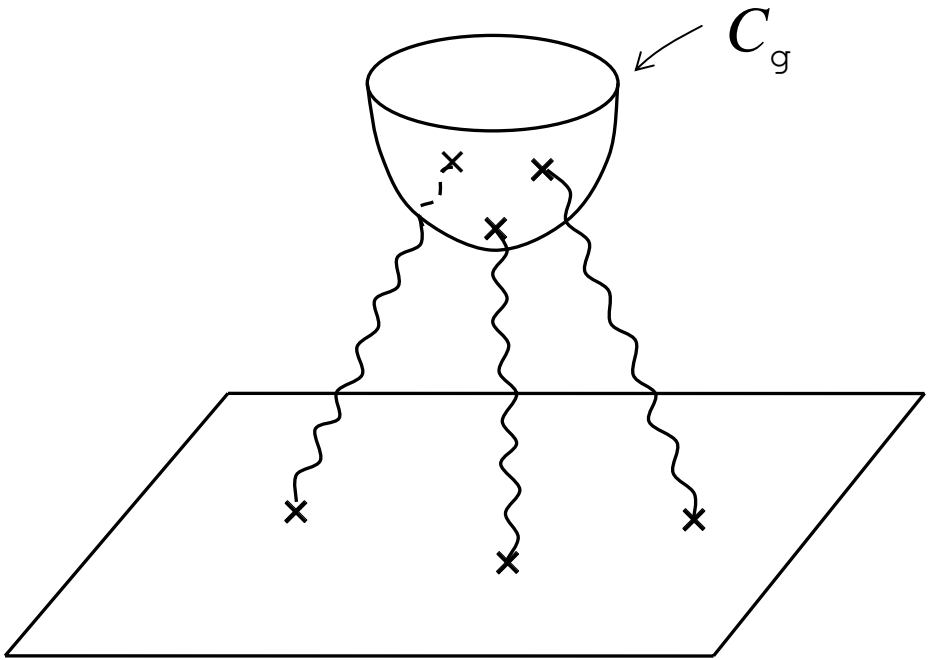}
\caption{A worldsheet far from D3-branes. The presence of the D3-branes are described by propagations 
of closed strings. }
   \label{gravity}
\end{figure}

On the other hand, we can perform another scale transformation which shrinks the system in the 
$x^\mu$-directions and expands in the $y^I$-directions, 
while keeping $C$ within the near-horizon region. 
Since this system is invariant under this transformation, 
the worldsheet path-integral for this transformed system should provide 
another way of evaluating the Wilson loop discussed above. 
We assume that the near-horizon region is large enough in order 
for the classical gravity approximation 
to be valid, which is the case if we take a large enough $\lambda$. 
Since the worldsheet is far away from the D3-branes in this case, the dominant contributions 
come from a worldsheet $\Sigma$ which has only one boundary $C_g$ related to $C$ by the scale 
transformation.  
The presence of the other boundaries lying on the D3-branes is now encoded by insertions on $\Sigma$ 
of vertex 
operators of massless closed string states, which describe the exchange of 
massless states between the D3-branes and $\Sigma$. 
The summation over worldsheets with various numbers of boundaries then corresponds to the summation of 
all possible massless state exchanges. 
As a result, $\Sigma$ feels that the space-time surrounding $\Sigma$ is 
curved, and therefore, the worldsheet path-integral is dominated by 
a worldsheet with the boundary $C_g$ 
which is the minimal surface in the near-horizon geometry of the D3-branes. 
See figure \ref{gravity}. 

We have argued that the scale invariance relates the Wilson loop in ${\cal N}=4$ SYM and the minimal 
surface in $AdS_5\times S^5$. 
It should be noted that we have shown an equivalence for the case $C_{\rm YM}\ne C_g$. 
In fact, the length scales $l$ of the loops are 
\begin{equation}
l(C_g)<l(C)<l(C_{\rm YM}). 
\end{equation}
However, since the Wilson loop and the area of the minimal surface, both defined appropriately, 
are independent of the length scale of the loop, 
the equivalence should also hold for the case $C_{\rm YM}=C_g$ which is 
the equivalence that $AdS$/CFT correspondence claims. 

In summary, we have shown that {\it the presence of the scale invariance implies $AdS$/CFT 
correspondence}. 
The important point of our argument is that the scale invariance exists at any intermediate situation, 
like the one in figure \ref{intermediate}, not only at the two limiting cases (figure \ref{gauge},
\ref{gravity}). 
It is also important that the quantities we have discussed are independent of the length scale, 
in order for the ordinary 
claim of $AdS$/CFT correspondence to hold. 
In section \ref{scale}, we will show that the desired symmetry really exists if we concentrate on the 
dynamics in the near-horizon region. 
As a result, the weakest version of $AdS$/CFT correspondence follows. 

\vspace{5mm}

\begin{figure}[tbp]
\includegraphics{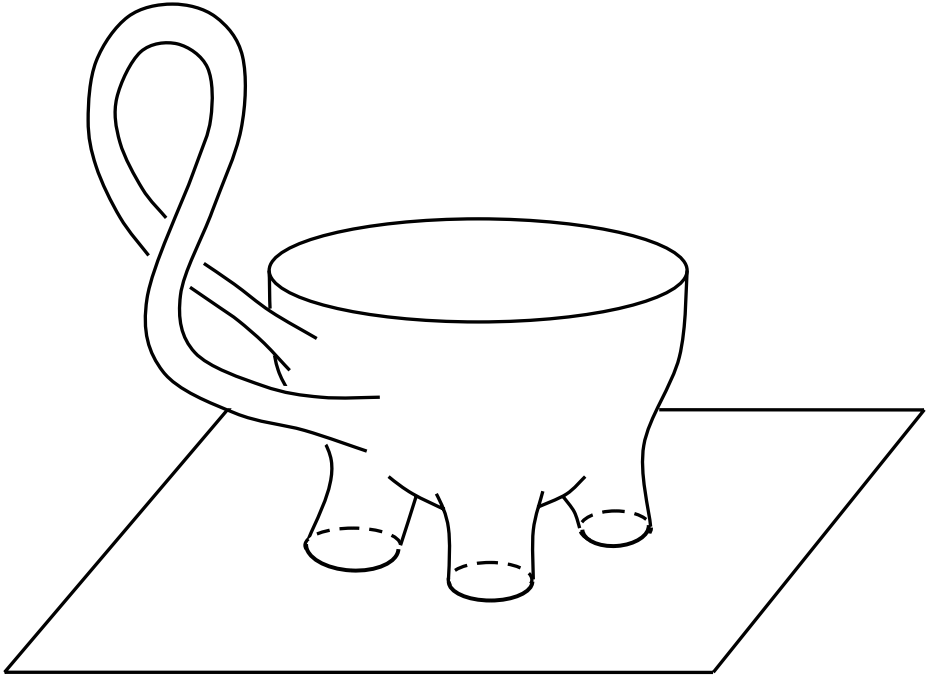}
\caption{A worldsheet with a handle which may be elongated beyond the limit of the near-horizon 
region. }
   \label{handle}
\end{figure}

It would be important to mention the effects of string loops to the above argument. 
Throughout this paper, we will not discuss worldsheets with handles. 
In addition to the technical difficulty of summing over string 
loop effects, there is a more serious issue. 
If we allow a handle on the worldsheet, then the handle may be elongated, as depicted in figure 
\ref{handle}. 
Since the handle includes the propagation of massless closed string states, the worldsheet may feel 
the space-time outside the near-horizon region. 
This implies that we cannot concentrate our attention on the near-horizon region appropriately, 
and the scale invariance is not available. 
In the following, we restrict ourselves to the standard large $N$ limit, that is, we take 
\begin{equation}
N \to \infty, \hspace{5mm} \lambda = g_sN = \mbox{fixed}. 
\end{equation}

To show the scale invariance, we 
will use the boundary state \cite{GG} of the D3-brane in the Green-Schwarz formalism in the light-cone 
gauge, 
which we review in the next section. 

\vspace{1cm}

\section{Boundary state}  \label{boundary}

\vspace{5mm}

In terms of closed strings, D-branes are described by boundary states. 
Since D-branes in Type IIB string 
preserve half of the 32 supercharges, the boundary state $|B\rangle$ should satisfy 
\begin{equation}
(Q+\beta^\perp\tilde{Q})_\alpha|B\rangle = 0, 
   \label{BPS}
\end{equation}
where $\beta^\perp$ is a product of gamma matrices. 
For a D3-brane extending along $x^{0,1,2,3}$-directions, 
\begin{equation}
\beta^\perp=\Gamma^0\Gamma^1\Gamma^2\Gamma^3
\end{equation}
is taken. 
We often denote the preserved supercharges as ${\cal Q}_\alpha$. 
Note that these are the supercharges of ${\cal N}=4$ SYM on the D3-branes. 

It will be convenient to employ the Green-Schwarz 
formalism which makes the space-time supersymmetry manifest. 
In addition, it is rather easy to handle R-R fields in this formalism. 
For notations and conventions, we basically follow \cite{GSW}. 
In the Green-Schwarz formalism, it is convenient 
to take the light-cone gauge to quantize the theory, but 
the Lorentz invariance is not manifest. 
In addition, the coordinate fields in the light-cone directions have to obey the Dirichlet boundary 
condition \cite{GG}, and therefore, the branes which are naturally realized as boundary states 
are instanton-like objects. 
In the following, we review the construction of the boundary state for a D3-brane \cite{GG}. 

For the boundary state of the D3-brane, the condition (\ref{BPS}) becomes  
\begin{eqnarray}
(Q^a+iM^{ab}\tilde{Q}^b)|B\rangle &=& 0, 
   \label{SUSY1} \\
(Q^{\dot{a}}+iM^{\dot{a}\dot{b}}\tilde{Q}^{\dot{b}})|B\rangle &=& 0, 
   \label{SUSY2} 
\end{eqnarray}
where we have decomposed the $SO(9,1)$ spinors into $SO(8)$ spinors, and we have defined 
\begin{equation}
M^{ab} = (\gamma^1\gamma^2\gamma^3\gamma^4)^{ab}, \hspace{5mm} 
M^{\dot{a}\dot{b}} = (\gamma^1\gamma^2\gamma^3\gamma^4)^{\dot{a}\dot{b}}. 
\end{equation}
There appears $i$ in front of $M^{ab}$ and $M^{\dot{a}\dot{b}}$ in (\ref{SUSY1})(\ref{SUSY2}) 
due to the double Wick rotation for obtaining the Euclidean 
D3-brane from the physical one. 

The D3-brane we consider is extended in $x^{1,2,3,4}$-directions, 
and localized in the other directions 
including the light-cone directions. 
This situation is realized by the condition 
\begin{equation}
(\alpha^i_n-M^{ij}\tilde{\alpha}_{-n}^j)|B\rangle = 0, \hspace{5mm} (n\ne0)
    \label{bosonic}
\end{equation}
where $\alpha_n^i$ and $\tilde{\alpha}_n^i$ are the oscillators of the bosonic coordinate fields 
$X^i$ with $i=1,\cdots,8$, and 
\begin{equation}
M^{ij} = \left[ 
\begin{array}{cc}
-I_{4\times 4} & 0 \\ 0 & I_{4\times4} 
\end{array}
\right]. 
\end{equation}
The condition for fermions $S^a$ and $\tilde{S}^a$ should be compatible with (\ref{bosonic}). 
That is, 
\begin{equation}
\left[{\cal Q}_\alpha\ ,\ \alpha_n^i-M^{ij}\tilde{\alpha}_{-n}^j\right]|B\rangle = 0 
   \label{compatible}
\end{equation}
should be imposed. 
Recall that the explicit forms of the supercharges are 
\begin{eqnarray}
Q^a &=& \sqrt{2p^+}S_0^a, \\
Q^{\dot{a}} &=& \frac1{\sqrt{p^+}}\gamma^{i\dot{a}a}\sum_{n\in{\bf Z}}S^a_{-n}\alpha^i_n. 
\end{eqnarray}
From (\ref{compatible}), we obtain 
\begin{equation}
(S^a_n+iM^{ab}\tilde{S}^b_{-n})|B\rangle = 0. 
   \label{fermionic}
\end{equation}
Note that the condition (\ref{fermionic}) for $n=0$ is nothing but the BPS condition (\ref{SUSY1}). 

The conditions (\ref{bosonic}) and (\ref{fermionic}) determines the dependence 
of the boundary state  on the non-zero modes as 
\begin{equation}
|B\rangle = \exp\left[ \sum_{n=1}^\infty\left(\frac1n M^{ij}\alpha_{-n}^i\tilde{\alpha}^j_{-n}
 -iM^{ab}S^a_{-n}\tilde{S}^b_{-n} \right)\right]|B_0\rangle. 
\end{equation}
The dependence of $|B_0\rangle$ on the zero modes of $S^a$ and $\tilde{S}^a$ 
is determined as follows. 
The lowest energy subspace of the Hilbert space of the left movers 
is spanned by the vector representation $|i\rangle$ 
and the conjugate spinor representation $|\dot{a}\rangle$, and similar for the right movers. 
$|B_0\rangle$ is thus constructed from a sum of tensor products of them. 
Since $|B\rangle$ is a source of the graviton, it is a bosonic state. 
We assume 
\begin{equation}
|B_0\rangle = \left( c^{ij}|i\rangle\tilde{|j\rangle}
 +c^{\dot{a}\dot{b}}|\dot{a}\rangle\tilde{|\dot{b}\rangle} \right) 
 |p^{1,2,3,4}=0\rangle |x^{5,6,7,8,+,-}=0\rangle 
 \left(\bigotimes_{n=1}^\infty |0_n\rangle \right), 
\end{equation}
where $|0_n\rangle$ is the Fock vacuum of the $n$-th mode. 
The condition (\ref{SUSY1}) implies 
\begin{eqnarray}
\gamma^{ia\dot{a}}c^{ij}-iM^{ab}\gamma^{jb\dot{b}}c^{\dot{a}\dot{b}} &=& 0, \\
\gamma^{ia\dot{a}}c^{\dot{a}\dot{b}}+iM^{ab}\gamma^{jb\dot{b}}c^{ij} &=& 0. 
\end{eqnarray}
It is easy to check that 
\begin{equation}
c^{ij} = CM^{ij}, \hspace{5mm} c^{\dot{a}\dot{b}} = -iCM^{\dot{a}\dot{b}},
\end{equation}
is a solution, where $C$ is a constant. 
We obtain 
\begin{equation}
|B_0\rangle = C\Bigl(M^{ij}|i\rangle\tilde{|j\rangle}
 -iM^{\dot{a}\dot{b}}|\dot{a}\rangle\tilde{|\dot{b}\rangle}\Bigr)
 |p^{1,2,3,4}=0\rangle |x^{5,6,7,8,+,-}=0\rangle 
 \left(\bigotimes_{n=1}^\infty |0_n\rangle \right). 
\end{equation}
It is straightforward to show that this state also satisfies (\ref{SUSY2}). 
The constant $C$ is determined to be unity by requiring the s-t duality.

\vspace{1cm}

\section{Scale transformation in worldsheet theory}   \label{scale}

\vspace{5mm}

We consider a perturbative description of Type IIB string with $N$ D3-branes in the flat background. 
The worldsheet action is that of free fields, 
\begin{equation}
S = -\frac1{\pi}\int d^2\sigma\left[ \frac12\partial_\alpha X^i\partial^\alpha X^i-iS^a\partial_-S^a
 -i\tilde{S}^a\partial_+\tilde{S}^a \right], 
   \label{WSaction}
\end{equation}
where we choose $\alpha'=\frac12$ in this section. 
The presence of the D3-branes is encoded by 
including contributions from worldsheets with boundaries. 
The back-reaction of the D3-branes to the background field configuration is taken into account 
by summing over 
all possible worldsheets with various number of boundaries. 
It is obvious that the worldsheet action (\ref{WSaction}) 
is not invariant under the scale transformation (\ref{isometry}). 
Therefore, the scale invariance is not manifest in the worldsheet 
description. 

In this section, we define the scale transformation of the worldsheet degrees of freedom, 
in terms of closed strings in a suitable 
manner, 
and first examine the transformation property of the 
boundary state constructed in the previous section. 
Next, we show that the presence of the D3-branes enables the scale transformation to be 
really a symmetry of the total system in the near-horizon region. 

\vspace{5mm}

\subsection{Definition of scale transformation}

\vspace{5mm}

We define the scale transformation of $X^i$ and their canonical momenta $P^i$ at $\tau=0$ as 
\begin{eqnarray}
\delta_\epsilon X^i(\sigma) &=& \epsilon M^{ij}X^j(\sigma), \\
\delta_\epsilon P^i(\sigma) &=& -\epsilon M^{ij}P^j(\sigma). 
   \label{P}
\end{eqnarray}
Note that we realize the transformation (\ref{isometry}) as a canonical transformation. 
Recall that the mode expansions of $X^i$ and $P^i$ at $\tau=0$ are 
\begin{eqnarray}
X^i(\sigma) &=& x^i+i\sqrt{\frac{\alpha'}2}\sum_{n\ne0}\frac1n(\alpha^i_n
 -\tilde{\alpha}^i_{-n})e^{-in\sigma}, \\
2\pi P^i(\sigma) &=& p^i+\sum_{n\ne0}(\alpha^i_n
 +\tilde{\alpha}^i_{-n})e^{-in\sigma}. 
\end{eqnarray}
These imply the scale transformation of the oscillators as 
\begin{eqnarray}
\delta_\epsilon \alpha^i_n &=& -\epsilon M^{ij}\tilde{\alpha}^j_{-n}, 
   \label{bosonic1} \\
\delta_\epsilon \tilde{\alpha}^i_n &=& -\epsilon M^{ij}\alpha^j_{-n}. 
   \label{bosonic2}
\end{eqnarray}
For later use, it should be noted that the light-cone momentum $p^+$ also transforms as 
\begin{equation}
\delta_\epsilon p^+ = -\epsilon p^+, 
   \label{p^+}
\end{equation}
since the light-cone coordinates are taken to obey the Dirichlet condition. 

The transformation of the ferminonic oscillators are determined by using the transformation of the 
supercharges 
\begin{equation}
\delta_\epsilon {\cal Q}_\alpha = \frac\epsilon2{\cal Q}_\alpha, 
   \label{algebra}
\end{equation}
which is derived from the transformation (\ref{P}) and the relation 
${\cal Q}^2\sim P^\mu$. 
Note that this is equivalent to the commutation relation between the supercharges and the dilatation 
operator in ${\cal N}=4$ SYM. 
Evaluating $\delta_\epsilon[{\cal Q}_\alpha,\alpha^i_n]$ and 
$\delta_\epsilon[{\cal Q}_\alpha,\tilde{\alpha}^i_n]$ in two ways, we obtain 
\begin{eqnarray}
\delta_\epsilon S^a_n &=& i\epsilon M^{ab}\tilde{S}^b_{-n}, \\
\delta_\epsilon \tilde{S}^a_n &=& -i\epsilon M^{ab}S^b_{-n},
\end{eqnarray}
where we have used the transformation property (\ref{p^+}). 
This transformation also preserves the commutation relations. 
Note that the symmetric property of $M^{ab}$ is crucial for the preservation of the commutation 
relations. 
The transformation of the fields $S^a(\sigma)$ and $\tilde{S}^a(\sigma)$ at $\tau=0$ is therefore 
\begin{eqnarray}
\delta_\epsilon S^a(\sigma) &=& i\epsilon M^{ab}\tilde{S}^b(\sigma), \\
\delta_\epsilon \tilde{S}^a(\sigma) &=& -i\epsilon M^{ab}S^b(\sigma).
\end{eqnarray}

Because of the presence of $i$, this transformation is not unitary. 
However, the appearance of $i$ here inherits the $i$ in (\ref{SUSY1})(\ref{SUSY2}), and therefore 
this should be regarded as an artifact for considering the Euclidean D-brane. 
If it were possible to construct a boundary state of the physical D-brane with manifest space-time 
supersymmetry, then the scale transformation of fermionic fields would also be an ordinary canonical 
transformation. 

\vspace{5mm}

\subsection{Boundary state}

\vspace{5mm}

In this subsection, we examine the transformation property of the boundary state $|B\rangle$. 
First, consider the transformation of non-zero mode oscillators in $|B\rangle$. 
Since each mode can be treated separately, we consider, for $n>0$, the transformation of 
\begin{equation}
|b_n\rangle = \exp\left[ \frac1nM^{ij}\alpha_{-n}^i\tilde{\alpha}_{-n}^j
 -iM^{ab}S_{-n}^a\tilde{S}_{-n}^b \right]|0_n\rangle. 
\end{equation}
It is important to notice that $|0_n\rangle$ is also transformed by the scale transformation, since 
it mixes the creation and annihilation operators as (\ref{bosonic1})(\ref{bosonic2}). 
That is, $\delta_\epsilon$ acts on $|0_n\rangle$ as a Bogoliubov transformation. 
Let $|0_n\rangle'$ be the transformed vacuum. 
The explicit form is 
\begin{equation}
|0_n\rangle' = \exp\left[ \frac{\epsilon}nM^{ij}\alpha_{-n}^i\tilde{\alpha}_{-n}^j
 -i\epsilon M^{ab}S_{-n}^a\tilde{S}_{-n}^b \right]|0_n\rangle. 
\end{equation}
The transformed state $|b_n\rangle'$ is therefore 
\begin{equation}
|b_n\rangle' = e^{A_n-\frac\epsilon nH_n}e^{\epsilon A_n}|0_n\rangle, 
\end{equation}
where 
\begin{eqnarray}
A_n &=& \frac1nM^{ij}\alpha^i_{-n}\tilde{\alpha}^j_{-n}-iM^{ab}S^a_{-n}\tilde{S}^b_{-n}, \\
H_n &=& \alpha^i_{-n}\alpha^i_n+\tilde{\alpha}^i_{-n}\tilde{\alpha}^i_n+nS^a_{-n}S^a_n
 +n\tilde{S}^a_{-n}\tilde{S}^a_n. 
\end{eqnarray}
Note that the constant terms, which could appear from the ordering of the oscillators, cancels each 
other by the supersymmetry. 
Using the commutation relation 
\begin{equation}
[H_n,A_n] = 2nA_n, 
\end{equation}
it is easy to show that 
\begin{eqnarray}
|b_n\rangle' &=& e^{A_n}e^{-\frac\epsilon nH_n}e^{-\epsilon A_n}e^{\epsilon A_n}|0_n\rangle 
  \nonumber \\
&=& e^{A_n}|0_n\rangle \nonumber \\
&=& |b_n\rangle. 
\end{eqnarray}

Next, let us consider the zero modes. 
The transformation of zero modes is  
\begin{eqnarray}
\delta_\epsilon S^a_0 &=& i\epsilon M^{ab}\tilde{S}^b_0, 
   \label{zero'} \\
\delta_\epsilon \tilde{S}^a_0 &=& -i\epsilon M^{ab}S^b_0, 
   \label{zero}
\end{eqnarray}
which is consistent with (\ref{algebra}). 
To examine the transformation of $|B_0\rangle$, let us construct the operator $D$ which generates the 
transformation (\ref{zero'})(\ref{zero}). 
The explicit form of $D$ is  
\begin{equation}
D = -M^{ab}S_0^a\tilde{S}^b_0+c{\bf 1}, 
\end{equation}
where the transformation is defined as $\delta_\epsilon S_0^a=i\epsilon[D,S^a_0]$ etc. 
Here $c$ is a constant which is not determined simply by the algebra, since we consider only the 
subgroup $SO(3,1)\times SO(1,1)$ of the $SO(4,2)$ in our worldsheet description. 

It is now straightforward to show that 
\begin{equation}
\delta_\epsilon |B_0\rangle = i\epsilon(-4i+c)|B_0\rangle. 
\end{equation}
Using the ambiguity of $c$, we can define the scale transformation so that $|B_0\rangle$ is invariant. 
Since the non-zero mode part of $|B\rangle$ is also invariant, it has been shown that $|B\rangle$ is 
scale invariant, for the above definition of the transformation. 

It is worth emphasizing again that the supersymmetry, as well as the symmetric property of $M^{ab}$, 
play crucial roles for 
the invariance of the boundary state under the scale transformation. 
If one considers a boundary state in, for example, bosonic string theory, then the variation of 
the boundary state is proportional to itself with a divergent coefficient. 
By a regularization of the divergence, one would obtain a non-vanishing term in the variation, even 
after subtracting a divergent terms.

\vspace{5mm}

\subsection{Variation of free energy}

\vspace{5mm}

We would like to evaluate the variation of the free energy of a string in the D3-brane background 
under the scale transformation. 
The variation for the case of figure \ref{intermediate} can be evaluated similarly, provided that 
we treat the 
transformation of the boundary $C$ properly. 

The free energy is defined as an infinite sum 
\begin{equation}
F(\lambda) = \sum_{n=0}^\infty \frac{F_n}{n!}\lambda^n, 
\end{equation}
where $F_n$ contains the contributions from worldsheets with $n$ boundaries. 
The scale transformation $\delta_\epsilon F(\lambda)$ of the free energy is thus determined by 
$\delta_\epsilon F_n$. 
The quantity $\delta_\epsilon F_n$ may have two contributions, 
one of which comes from the variation of the 
action, the other of which comes from the variation of the boundary. 
As shown in the previous section, each boundary is invariant under the scale transformation. 
Therefore, we only need to consider the variation of the action. 
Since the scale transformation is defined on the oscillators, it is convenient to consider in the 
canonical formalism. 
The path-integral measure is 
\begin{equation}
\int {\cal D}X{\cal D}P{\cal D}S{\cal D}\tilde{S}\ e^{iS_B+iS_F}, 
\end{equation}
where 
\begin{eqnarray}
S_B &=& \int d^2\sigma \left[ P^i\partial_\tau X^i-\left( \frac{\pi}2(P^i)^2+\frac1{2\pi}
 (\partial_\sigma X^i)^2 \right) \right], \\
S_F &=& \int d^2\sigma \left[ \frac i{2\pi}(S^a\partial_\tau S^a+\tilde{S}^a\partial_\tau \tilde{S}^a)
 -\frac i{2\pi}(S^a\partial_\sigma S^a-\tilde{S}^a\partial_\sigma \tilde{S}^a) \right]. 
\end{eqnarray}
Under the scale transformation, $S_B$ and $S_F$ transform as 
\begin{eqnarray}
\delta_\epsilon S_B &=& \int d^2\sigma\left[ -\pi\epsilon M^{ij}P^iP^j+\frac{\epsilon}\pi M^{ij}
 \partial_\sigma X^i\partial_\sigma X^j \right], \\
\delta_\epsilon S_F &=& \int d^2\sigma\ \frac\epsilon\pi M^{ab}S^a\partial_\sigma\tilde{S}^b. 
\end{eqnarray}
Performing the path-integral of $P^i$, we can deduce the transformation of the fields in the 
Lagrangian formalism. 
It can be shown that the transformation is nothing but 
\begin{eqnarray}
\delta_\epsilon X^i &=& \epsilon M^{ij}X^j, \\
\delta_\epsilon S^a &=& i\epsilon M^{ab}\tilde{S}^a, \\
\delta_\epsilon \tilde{S}^a &=& -i\epsilon M^{ab}S^a, 
\end{eqnarray}
and the variation of the action is therefore 
\begin{equation}
\delta_\epsilon S = \frac\epsilon\pi\int d^2\sigma \left[ -M^{ij}\partial_\alpha X^i\partial^\alpha
 X^i+M^{ab}S^a\partial_\sigma\tilde{S}^a \right]. 
   \label{deltaS}
\end{equation}
The first term of the RHS can be regarded as a vertex operator of a massless graviton with zero 
momentum. 
Note that this vertex operator does not contain the dilaton contribution, since $M^{ij}$ is traceless, 
which is suitable for later discussion. 
The transformation of the supercharge (\ref{algebra}) 
implies that $\delta_\epsilon S$ is invariant under 
${\cal Q}_\alpha$, and therefore, the second term of the RHS of (\ref{deltaS}) should be related to 
the vertex operator of a R-R field. 

It has been shown that the variation of the worldsheet action provides a linear combination 
of vertex operators of massless fields, which are invariant under ${\cal Q}_\alpha$. 
Recall that the bosonic state in the massless sector of the Hilbert space which is invariant under 
${\cal Q}_\alpha$ is proportional to $|B_0\rangle$. 
Therefore, the variation $\delta_\epsilon S$ of the action can be replaced with the insertion of the 
state $|B_0\rangle$. 
Schematically, we denote 
\begin{equation}
\delta_\epsilon F_n = \epsilon F_n(|B_0\rangle). 
\end{equation}
The tracelessness of $M^{ij}$ matches with the fact that the D3-brane is not a source of the dilaton. 

\begin{figure}[tbp]
\includegraphics{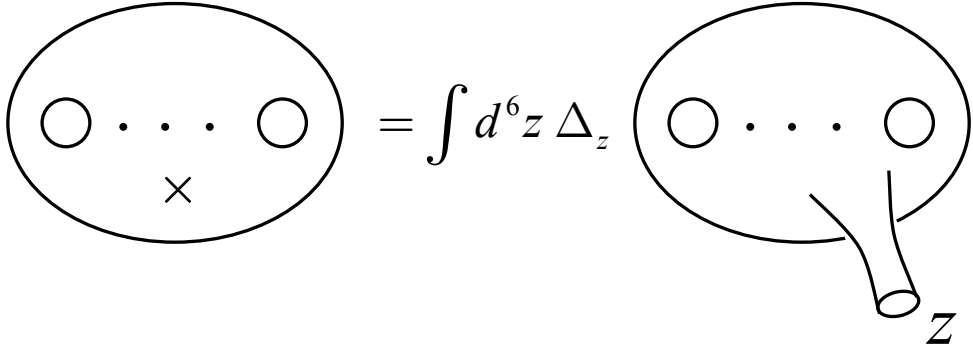}
\caption{LSZ-like reduction. The cross represents the insertion of $\delta_\epsilon S$ or 
$|B_0\rangle$.}
   \label{LSZ-like}
\end{figure}

To evaluate $F_n(|B_0\rangle)$, consider a worldsheet path-integral $F_{n+1}(z)$ with $n+1$ 
boundaries, one of whose boundary is placed at $x^i=z^i$ for $i=5,\cdots,10$. 
Note that the D3-branes we have discussed so far are placed at $x^i=0$. 
In other words, we place another set of D3-branes which are parallel to the original branes but their 
positions are different. 
It is possible to obtain $F_n(|B_0\rangle)$ from $F_{n+1}(z)$ through an LSZ-like procedure: 
\begin{equation}
F_n(|B_0\rangle) = \int d^6z\ \Delta_zF_{n+1}(z),  
   \label{LSZ}
\end{equation}
where $\Delta_z$ is the Laplacian on ${\bf R}^6$. 
See figure \ref{LSZ-like} for an image of this procedure. 
The variation of the total free energy is therefore 
\begin{equation}
\delta_\epsilon F(\lambda) = \epsilon \int d^6z\ \Delta_zF(\lambda,z), 
   \label{totalvariation}
\end{equation}
where 
\begin{equation}
F(\lambda,z) = \sum_{n=0}^\infty \frac{\lambda^n}{n!}F_{n+1}(z). 
\end{equation}

In the following, we will focus on a situation in which some operators placed around $|x|=y$ are 
inserted to the worldsheet. 
One example of such situations is to put a circular source of string, as 
in figure \ref{intermediate}. 
In this situation, the worldsheet is stretched in the region $0\le |x|\le y$. 
If we take $|z|$ to be larger than $y$, then the corresponding worldsheet has a thin tube connecting 
the boundary and the body of the worldsheet. 
The tube represents the propagation of closed string states, and the dominant contribution comes from 
the massless propagation. 
As a result, $F(\lambda,z)$ behaves as $|z|^{-4}$ for large $|z|$. 
For the range $0\le |z|\le y$, we assume that $F(\lambda,z)$ varies slowly with $|z|$. 

Let us calculate the integral 
\begin{equation}
I = \int d^6z\ \Delta_z f(z), 
\end{equation}
where 
\begin{equation}
f(z) = \left\{ 
\begin{array}{cc}
f(0), & (0\le |z|\le y) \\ [3mm] \displaystyle{\frac{y^4}{|z|^4}f(0)}, & (|z|>y)
\end{array}
\right.
\end{equation}
is a typical example of the function having the property assumed for $F(\lambda,z)$. 
It is easy to check that $I=-4\ vol(S^5)y^4f(0)$. 
Similarly, the RHS of (\ref{totalvariation}) would be estimated as 
\begin{equation}
\int d^6z \ \Delta_z F(\lambda,z) \sim y^4C(\lambda,y)F(\lambda,z=0), 
   \label{estimate}
\end{equation}
where $C(\lambda,y)$ is assumed to be of order one. 
Noticing that $F_{n+1}(z=0)=F_{n+1}$, we obtain 
\begin{eqnarray}
\delta_\epsilon F(\lambda) 
&\sim& \epsilon y^4C(\lambda,y)\sum_{n=0}^\infty \frac{\lambda^n}{n!}F_{n+1} \nonumber \\
&=& \epsilon y^4C(\lambda,y)\partial_\lambda F(\lambda). 
   \label{variation}
\end{eqnarray}
In other words, the free energy transforms as 
\begin{equation}
F(\lambda) \to F(\lambda+\epsilon y^4C(\lambda,y)), 
\end{equation}
which indicates {\it the existence of the 
scale invariance in the near-horizon region $y^4<<\lambda$}. 
Note that the sum of the infinite number of worldsheets with 
boundaries is crucial for the existence of this scale 
invariance. 

The discussion so far is based on a power series expansion in terms of $\lambda$, and it is not 
obvious whether the transformation property (\ref{variation}) also holds for large $\lambda$. 
To check the validity of (\ref{variation}) for large $\lambda$, 
let us consider an effective sigma model description of this system. 
Since $\lambda$ is large, we can take $y$ large enough so that the classical gravity approximation 
is valid (see figure \ref{gravity}), and this fact verifies the use of the sigma model. 
The presence of the D3-branes is realized in the sigma model by a non-trivial background field 
configuration, and the free energy of the model would be a functional of the background fields. 
Let us focus only on the metric, for simplicity. 
Then, due to (\ref{deltaG}), the free energy $F[g]$ satisfies
\begin{equation}
\delta_\epsilon F[g] \sim \epsilon \frac{y^4}{\pi}\partial_\lambda F[g], 
   \label{sigmamodel}
\end{equation}
which is the same relation with (\ref{variation}) up to a factor of order one. 
This strongly suggests that the free energy defined as a power series of $\lambda$ can be continued 
analytically to the region of large $\lambda$ where the gravity description is valid, and therefore, 
the scale invariance in the near-horizon region would exist for any $\lambda$.  

\vspace{5mm}

So far, we have only considered the infinitesimal transformation. 
A remarkable property of the scale transformation is that it brings the original free theory into 
{\it another free theory} which is obtained via a field redefinition from 
the original one, as shown in (\ref{deltaS}). 
The boundary state $|B\rangle$ has the same form for both the original and the redefined field 
variables. 
Therefore, we can perform the 
transformation repeatedly to obtain a finite transformation. 
The finite transformation of the worldsheet fields is 
\begin{eqnarray}
X^i(\sigma) &\to& e^{M^{ii}\theta}X^i(\sigma), \\
S^a(\sigma) &\to& \cos\theta S^a(\sigma)+i\sin\theta M^{ab}\tilde{S}^b(\sigma), \\
\tilde{S}^a(\sigma) &\to& -i\sin\theta M^{ab}S^b(\sigma)+\cos\theta \tilde{S}^a. 
\end{eqnarray}
It is this fact that 
enables us to take the worldsheet of figure \ref{gauge} to the one of figure \ref{gravity} 
and vice versa.

\vspace{1cm}

\section{Supersymmetric Wilson loop}   \label{Zarembo}

\vspace{5mm}

In this section, we show that the functional 
form of the supersymmetric Wilson loop 
is naturally derived from our worldsheet point of view. 

\begin{figure}[tbp]
\includegraphics{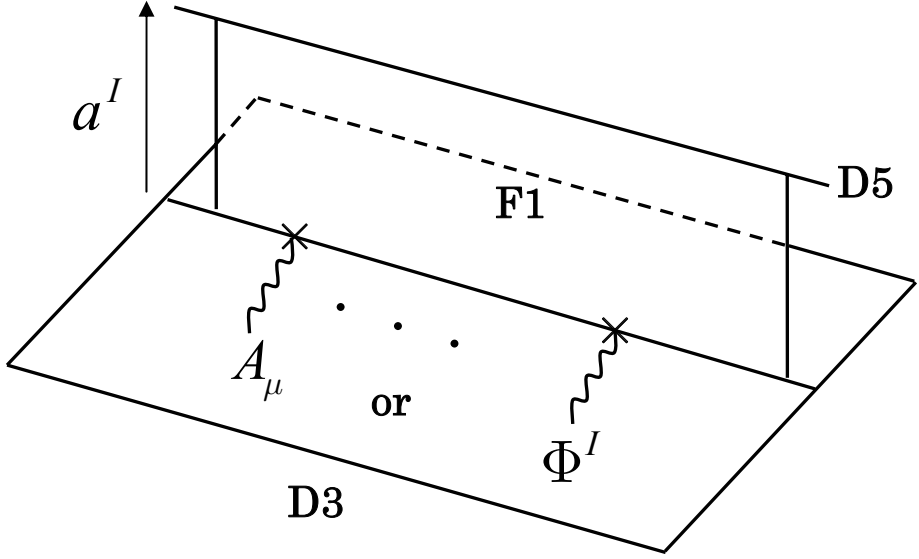}
\caption{A worldsheet configuration for a supersymmetric Wilson loop. The string is stretched between 
D3-branes and D5-brane. Open string states are emitted and absorbed by the string. }
   \label{D3-D5}
\end{figure}

A Wilson loop on the D3-branes was constructed, as mentioned in section \ref{sketch}, 
by putting a source of 
string in a target space-time, and by stretching a string between them. 
For definiteness, we choose a D5-brane as the source, which does not have common longitudinal spatial 
directions with the D3-branes. 
The D-branes extend in the directions indicated as follows. 

\vspace{5mm}

\begin{center}
\begin{tabular}{c|cccccccccc}
\hline 
 & 0 & 1 & 2 & 3 & 4 & 5 & 6 & 7 & 8 & 9 \\
\hline
D3 & $\bigcirc$ & $\bigcirc$ & $\bigcirc$ & $\bigcirc$ & & & & & & \\
D5 & $\bigcirc$ & & & & & $\bigcirc$ & $\bigcirc$ & $\bigcirc$ & $\bigcirc$ & $\bigcirc$ \\
F1 & $\bigcirc$ & & & & $\bigcirc$ & & & & & \\
\hline
\end{tabular}
\end{center}

\vspace{5mm}

This brane configuration preserves 8 supercharges, which we denote as 
$\xi^\alpha Q_\alpha+\tilde{\xi}^\alpha \tilde{Q}_\alpha$ where $\xi$ and $\tilde{\xi}$ are 
restricted as 
\begin{eqnarray}
\xi &=& \Gamma^0\Gamma^1\Gamma^2\Gamma^3\tilde{\xi}, \\
\xi &=& \Gamma^0\Gamma^5\Gamma^6\Gamma^7\Gamma^8\Gamma^9\tilde{\xi}. 
\end{eqnarray}
Recalling that both $\xi$ and $\tilde{\xi}$ are spinors with, say, positive chirality, of $SO(9,1)$, 
these supercharges are preserved even after 
a string is stretched between the D3-branes and the D5-brane, as depicted in figure \ref{D3-D5}. 
Therefore, the Wilson loop constructed here should preserve half the supersymmetry of ${\cal N}=4$ 
SYM. 
Its functional form is 
\begin{equation}
W(C) = \frac1N\mbox{Tr}\ \exp\left[ i\int_C ds ( \dot{x}^\mu A_\mu+|\dot{x}|\theta^I\Phi^I) \right]. 
   \label{WZ}
\end{equation}
This construction of the Wilson loop was also discussed in \cite{Yamaguchi}\cite{Gomis}. 

We would like to derive (\ref{WZ}). 
The coupling of the gauge fields and the scalars on the D3-branes to the worldsheet is described by 
the corresponding vertex operators 
\begin{equation}
\xi_\mu(p)\partial_\tau X^\mu e^{ipX}, \hspace{5mm} 
\varphi_I(p)\partial_\sigma X^I e^{ipX}, 
\end{equation}
respectively. 
Recall that the mode expansion of the open string stretched between the D3-branes and the D5-brane 
is 
\begin{eqnarray}
X^\mu &=& x^\mu+2\alpha'p^\mu\tau+\cdots, \\
X^I &=& \frac{a^I}\pi\sigma+\cdots, 
\end{eqnarray}
where $a^I$ is a vector indicating the separation of the D3-branes and the D5-brane, and 
the dots indicate the oscillation parts of the expansion. 
For our choice of $C$, only the $p^0$ component is non-zero. 
The Virasoro constraint implies 
\begin{equation}
2\alpha'p^0 = \frac1\pi|a^I|. 
\end{equation}
This relation leads 
\begin{equation}
\xi_\mu(p)\partial_\tau X^\mu+\varphi_I(p)\partial_\sigma X^I 
= 2\alpha'\left( p^0\xi_0(p)+p^0\theta^I\varphi_i(p) \right) + \cdots, 
\end{equation}
where $\theta^I$ is a unit vector parallel to $a^I$. 
Therefore, the coupling induced by (\ref{WZ}), including the relative coefficient, 
is correctly reproduced from the worldsheet picture.

\vspace{1cm}

\section{Discussion}    \label{discuss}

\vspace{5mm}

We have shown that there is a scale invariance in the worldsheet theory of Type IIB string with 
D3-branes as far as closed string loops are suppressed, if we only consider in the near-horizon 
region. 
This scale invariance actually implies the $AdS$/CFT correspondence for the Wilson loops in 
${\cal N}=4$ SYM on the D3-branes and the minimal surfaces in the corresponding $AdS_5\times S^5$ 
space-time. 
In the gauge theory point of view, we take the large $N$ limit, and we choose a large 
't Hooft coupling when we compare it with the results from classical gravity. 
Therefore, we have shown a part of the weakest version of the $AdS$/CFT correspondence. 
Our point of view also provides a natural understanding of the functional form of the supersymmetric 
Wilson 
loop operator in terms of the worldsheet stretched between the D3-branes and a D5-brane. 

It is desired to examine whether it is possible to argue a more stronger version of $AdS$/CFT 
correspondence in the worldsheet point of view. 
Since this is a perturbative description, the string coupling must be taken to be small. 
As discussed at the end of section \ref{sketch}, this is also crucial for considering the near-horizon 
region appropriately. 
The number $N$ of the D3-branes is not necessarily large in the worldsheet description. 
However, if $N$ is finite, then $\lambda$ must be small, and the near-horizon region is also small, 
as shown in section \ref{scale}. 
If the width of the near-horizon region is of order of the string scale, then worldsheet may fluctuate 
and reach the outside of the near-horizon region, and the scale invariance will not be available. 
It seems natural to think that the ordinary claims of $AdS$/CFT might be modified in this situation. 
It is possible to take a large $N$ limit, and obtain a {\it finite} $\lambda$ which is bigger than the 
string scale, but not big enough for the classical gravity description to be valid. 
In this case, it would be necessary to include $\alpha'$-corrections, 
and our point of view may provide a guideline for how to include the corrections. 
The strongest version, that is, both $N$ and $\lambda$ are finite, is beyond the reach of our point 
of view, and a non-perturbative formulation of string theory will be necessary to discuss this issue. 
Of course, there might be a possibility of occurring 
miraculous cancellations among worldsheets due to supersymmetry, 
and as a result, the scale invariance might be realized more generally. 
The range of validity of $AdS$/CFT correspondence in a stronger version was also discussed in 
\cite{Park}. 

\begin{figure}[tbp]
\includegraphics{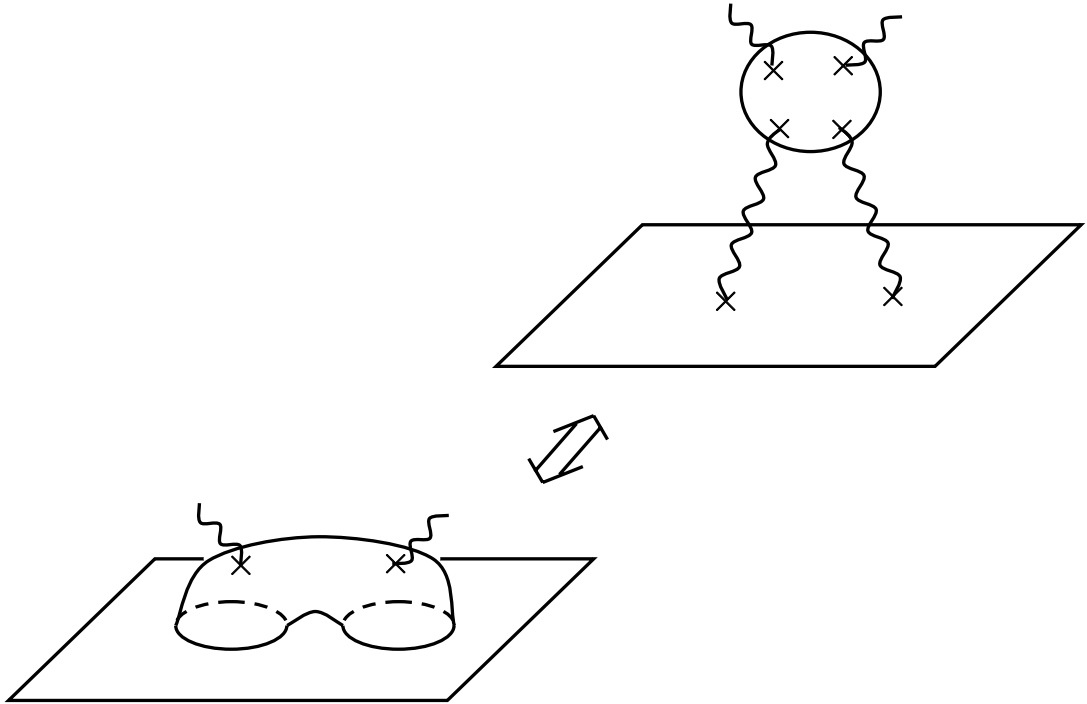}
\caption{GKPW relation. }
   \label{GKPW}
\end{figure}

It might be possible to apply the line of our arguments to less symmetric cases. 
It seems that one of the significant differences from the ${\cal N}=4$ version will be the 
transformation property of boundary states. 
To illustrate the difference, let us consider a boundary state in bosonic string theory 
\begin{equation}
|B\rangle = \exp\left[ \sum_{n=1}^\infty\frac1nM^{ij}\alpha_{-n}^i\tilde{\alpha}_{-n}^j \right]
 |0\rangle, 
\end{equation}
where $i$ and $j$ run from 1 to 24, and $M^{ij}$ is suitably defined. 
The transformation of the oscillators will be the same as (\ref{bosonic1})(\ref{bosonic2}). 
A naive calculation shows that the transformed state $|B\rangle'$ 
is proportional to $\exp[-24\epsilon \sum_{n=1}^\infty 1]$, and therefore some regularization is 
necessary. 
We employ the regularization 
\begin{equation}
|B\rangle \to e^{-\epsilon'H}|B\rangle, 
\end{equation}
where $H$ is the Hamiltonian, and take $\epsilon'\to0$ limit at the end of calculations. 
The result is 
\begin{equation}
\delta_\epsilon |B\rangle = \frac{12}{\epsilon'}|B\rangle-12(H+1)|B\rangle. 
\end{equation}
The finite part of the RHS has an interesting form, since this would be rewritten as a derivative 
of the boundary state with respect to a moduli parameter. 
This would imply that the boundary of the moduli space of Riemann surfaces with boundaries would 
be relevant for evaluation of $\delta_\epsilon F(\lambda)$. 
This might be important when one considers a case with a generic D-branes.

\begin{figure}[tbp]
\includegraphics{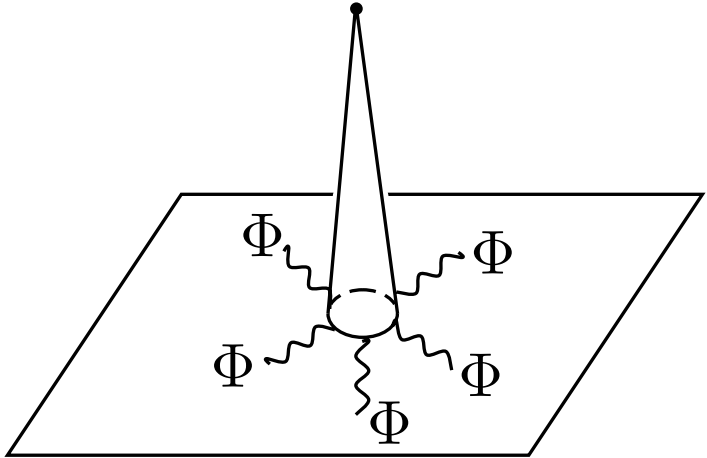}
\caption{Insertion of a BMN operator. }
   \label{BMN}
\end{figure}

In our argument, the only possible subtle point would be the estimate (\ref{estimate}). 
Although we could not completely verify the validity of (\ref{estimate}), we think 
it is already a 
great achievement that we could reduce $AdS$/CFT correspondence to a concrete problem of evaluating 
the LSH of (\ref{estimate}). 
We expect that our approach presented in this paper will provide a way of proving $AdS$/CFT 
correspondence, since it is now reduced to a problem in a free theory. 

It would be plausible if our argument can be extended to apply 
to the examination of the GKPW relation \cite{GKP}\cite{W}. 
Naively, it should be possible, since the Wilson loop is a source of generic string states, and 
therefore, the consideration of multiple Wilson loops would implies the GKPW relation. 
An intuitive image of how to show the GKPW relation is depicted in figure \ref{GKPW}. 
To describe 
an insertion of a BMN operator \cite{BMN} on the D3-branes, for example, it would be suitable 
to choose a Euclidean D3-brane which is perpendicular to the original D3-branes as a source of string. 
See figure \ref{BMN}. 
We hope to report on these issues elsewhere. 

\vspace{2cm}

{\bf \large Acknowledgments}

\vspace{5mm}

We would like to thank T.Azeyanagi, Y.Matsuo for valuable discussions. 
This work is supported by the Grant-
in-Aid for the 21st Century COE "Center 
for Diversity and Universality in Physics" from the Ministry of Education, 
Culture, Sports, Science and Technology (MEXT) of Japan.
The research of T.S is supported in part by JSPS Research Fellowships for Young Scientists.

\end{document}